# Robust predictive model for Carriers, Infections and Recoveries (CIR): predicting death rates for CoVid-19 in Spain

Efrén M. Benavides

Universidad Politécnica de Madrid

March 30, 2020

## I. Introduction

Since the end of January, CoVid-19 has spread through Europe generating an unprecedented medical collapse, firstly in Italy, secondly in Spain. Medical staff and services (especially Intensive Care Units – ICUs), mortuary services, sanitary suppliers, etc. appear all overwhelmed by the number of new patients and deaths per day: current predictive models seem to have underestimated the rate of infection.

This article presents a new model to predict the evolution of infectious diseases under uncertainty or low-quality information, just as it has happened in the initial scenario during the spread of CoVid-19 in China and Europe. The model uses a low number of input parameters and stochastic distributions. It is expected that this will provide the model with the required robustness to accurately predict demand on services. In particular, the model has been used to predict deceases, but it can be easily modified to predict the demand of ICUs or mechanical ventilators under different restraint policies. To achieve this goal, the model implements the following four key characteristics:

1. It keeps track of the date of infection of a single individual.
2. It uses stochastic distributions to aggregate individuals who share the same date of infection.
3. It uses two types of infections: mild and serious (this can be modified or extended).
4. It keeps track of the number of **C**arriers (instead of the number of infectious), **I**nfections (instead of the number of susceptible) and **R**ecoveries (instead of the number of recovered).

For the previous reasons, the mathematical structure of the proposed model, a Carriers-Infections-Recoveries (CIR) model, departs significantly from the one of a Susceptible-Infectious-Recovered (SIR) model.

The kernel of the presented model consists of four differential equations: two for Infections and two for Recoveries. The kernel can be complemented with additional differential equations, one for each sanitary event to be predicted. In the case presented in this article two additional differential equations are implemented: one for predicting the input rate of medical services, say 'hospitalization'; the other for predicting the output rate, which includes recovery and death.

This new set of six differential equations has been used to predict the death rate in Spain, demonstrating today an excellent agreement with the actual numbers. One important fact is that only one set of input parameters has been used for the prediction: this means that the input parameters are the same since the onset of the infection and hence no readjustment of the parameters has been required up to date to fit the results. For the case studied and at the moment of this publication, it seems that the proposed model has enough predictive accuracy.

## II. Predictive model: Infections and Recoveries

Let $h(\Delta t, \mu_R)dt$ be the probability that an infected individual has of stopping being contagious just in time $t$ after living with the virus during a period of time $\Delta t$; being $\mu_R$ a parameter distribution which is discussed in Appendix A (note that, if required, the distribution could accommodate more parameters or be changed by other ones). Here, 'stopping being contagious' means to reach immunity or death, which in both cases produces a barrier to propagation. According to the discussion in Appendix B, I suppose that there are two groups of contagious individuals: mild (type 1) and serious (type 2). The difference is that type 2 can recover or die whereas type 1 can only recover. Each type follows a different probability distribution, $h(\Delta t, \mu_{Ri})$, where $i \in \{1,2\}$ identifies the type (note that, if necessary, the model can easily accommodate more types).

Let $dI_i(t) = f_i(t)dt$ be the number of new infections of type $i \in \{1,2\}$ at time t during the period of time $dt$. It is convenient to write this equation in a past time $t_0 \in [0,t]$ and during a past period $dt_0$

$$dI_i(t_0) = f_i(t_0)dt_0 \qquad (1)$$

One of the main novelties of the model comes from assuming that infected people have to spend a different period of time to become recovered (immune or dead), and for this reason they are labelled with the date of infection $t_0$. In effect, assume that $dI_i(t_0)$ is the number of people who were infected just in the instant $t_0$ and that such people have lived with the illness for a period of time $\Delta t$ to reach the current date, then the number of such people who become recovered is $h(\Delta t, \mu_{Ri})dI_i(t_0)dt$. As long as $t_0$ can be any past time, the total number of new recovered people at time $t$ is $dt \int_0^t h(\Delta t, \mu_{Ri})dI_i(t_0)$. Let $dR_i(t)$ be such number, i.e., the total number of new recoveries (immunity or death) generated from type $i$ at time $t$ during the period $dt$, then

$$dR_i(t) = dt \int_0^t h(\Delta t, \mu_{Ri})dI_i(t_0)$$

Using (1) and that the moment of contagion $t_0$ is related to the current moment $t$ through the equation $t_0 = t - \Delta t$, last equation yields to

$$dR_i(t) = dt \int_0^t f_i(t_0)h(\Delta t, \mu_{Ri})dt_0 = dt \int_0^t f_i(t - \Delta t)h(\Delta t, \mu_{Ri})d\Delta t$$

Considering that eq. (1) implies $\frac{dI_i}{dt}(t) = f_i(t)$, it is $f_i(t - \Delta t) = \frac{dI_i}{dt}(t - \Delta t)$, and last equation looks as

$$dR_i(t) = dt \int_0^t \frac{dI_i(t - \Delta t)}{dt} h(\Delta t, \mu_{Ri})d\Delta t \qquad (2)$$

Equations (1) and (2) let find $I_i$ and $R_i$ with $i \in \{1,2\}$ for a set of initial conditions: in this paper it is supposed that the infection starts with one mild infection, that is, with the following initial conditions: $I_1(0) = 1$, $I_2(0) = 0$ and $R_i(0) = 0$.

It is remarkable that, in this model, $I_i$ and $R_i$ are respectively the total number of infections and the total number of recoveries which comes from each type. By definition, the carriers (contagious people) are obtained by subtracting the recoveries from the infections. Thus, calling $C_i(t)$ to the number of carriers at time $t$ of type $i$, we have that

$$C_i(t) = I_i(t) - R_i(t) \qquad (3)$$

Note that equations (1) to (3) make a huge difference with respect to the SIR model. This is because the convolution in eq. (2), which comes from keeping track of the date of infection, obligates to separate the number of contagious individuals from the number of infections and, hence, the number of carriers substitutes the SIR's number of infectious. For this reason, the present model is a CIR model where

the total number of carriers is

$$C(t) = C_1(t) + C_2(t) \qquad (4)$$

the total number of infections is

$$I(t) = I_1(t) + I_2(t) \qquad (5)$$

and the total number of recoveries is

$$R(t) = R_1(t) + R_2(t) \qquad (6)$$

**Contagion model**

The number of new infections is proportional to the number of free carriers, who are those carriers whose mobility has not been restricted: $C(t) - E(t)$. Here $E(t)$ is the number of carriers who are isolated in hospitals or who keep themselves at home. I suppose that they are a fraction $\alpha$ of the serious (type 2) carriers: $E(t) = \alpha C_2(t)$. The number of new infections is also proportional to the frequency $\omega_i(t)$ that a carrier has of finding people who are susceptible to being infected as type $i$. Therefore, finally, the rate of infections is given by

$$f_i(t) = \omega_i(t)[C(t) - \alpha C_2(t)] = \omega_i(t)[C_1(t) + (1-\alpha)C_2(t)] \qquad (7)$$

**Propagation model**

Function $\omega_i(t)$ includes (a) the average frequency $\omega$, which is the average number of persons who an average person finds per day; (b) the factor $\gamma$, which measures the average success of contagion; (c) the factor $\phi_i$, which is the average fraction of infections that will be of type $i \in \{1,2\}$ where $\phi_2 = 1 - \phi_1 = \phi_r$ (risk fraction); and (d) the fraction of susceptible people.

Let $P$ be the total susceptible (available) population. Since carriers and recoveries are not susceptible to being infected, the susceptible people are $S(t) = P - C(t) - R(t)$ and, using eq. (3), this becomes $S(t) = P - I(t)$. Thus, the fraction of susceptible people is $1 - I(t)/P$.

Previous reflections let us write

$$\omega_i(t) = \left(1 - \frac{I(t)}{P}\right)\omega\gamma\phi_i \qquad (8)$$

## III. Differential equations

It is convenient to define the following fractions

$$\phi_{Ci}(t) = \frac{C_i(t)}{P} \tag{9}$$

$$\phi_{Ii}(t) = \frac{I_i(t)}{P} \tag{10}$$

$$\phi_{Ri}(t) = \frac{R_i(t)}{P} \tag{11}$$

Using these fractions and placing equations (7) and (8) into (1) and (2) we reach the four differential equations that determine the temporal evolution of the infections and recoveries of each type $i$:

$$\frac{d\phi_{Ii}(t)}{dt} = [1 - \phi_{I1}(t) - \phi_{I2}(t)][\phi_{C1}(t) + (1-\alpha)\phi_{C2}(t)]\omega\gamma\phi_i \tag{12}$$

$$\frac{d\phi_{Ri}(t)}{dt} = \int_0^t \frac{d\phi_{Ii}(t-\Delta t)}{dt} h(\Delta t, \mu_{Ri}) d\Delta t \tag{13}$$

This system of four differential equations is highly non-linear and very different from the SIR models used normally. Appendix C shows that, for small fractions and only one type of contagious people, it has a stationary solution when time tends to infinity.

## IV. Hospitalization and death

Type-2 carriers are seriously infected people who, after the incubation period, will use medical services, say hospitals, UCIs, etc. These 'hospitalized' patients will leave the medical services, after a period of time, because they have died or because they have recovered enough.

Let $h(\Delta t, \mu_{IH})dt$ be the probability that an infected individual has of being an input in a medical service just in time $t$ after living with the virus during a period of time $\Delta t$. Let $h(\Delta t, \mu_{OH})dt$ be the probability that an infected individual has of being an output in a medical service just in time $t$ after being in the hospital a period of time $\Delta t$. Both, $\mu_{IH}$ and $\mu_{OH}$, are parameters which are discussed in Appendix A. Therefore, following a similar argumentation to the one given for equation (2), the fractions of inputs and outputs are calculated by the following two differential equations

$$\frac{d\phi_{IH}(t)}{dt} = \int_0^t h(\Delta t, \mu_{IH}) d\phi_{C2}(t_0) = \int_0^t \frac{d\phi_{I2}(t-\Delta t)}{dt} h(\Delta t, \mu_{IH}) d\Delta t \tag{14}$$

$$\frac{d\phi_{OH}(t)}{dt} = \int_0^t h(\Delta t, \mu_{OH}) d\phi_{IH}(t_0) = \int_0^t \frac{d\phi_{IH}(t-\Delta t)}{dt} h(\Delta t, \mu_{OH}) d\Delta t \tag{15}$$

$$\phi_H(t) = \phi_{IH}(t) - \phi_{OH}(t) \tag{16}$$

Equation (16) lets calculate the number of people $\phi_H(t)P$ using the medical service at any time. In addition, it is supposed that deaths are a constant fraction $t_D$ of the people leaving the hospital, so that

$$\phi_D(t) = t_D \phi_{OH}(t) \qquad (17)$$

**Numerical calculation**

The set of six differential equations given by (12) to (15) (as well as the convolution integrals inside them) have been numerically integrated with a low order integrator (Euler) using a time step of 0.4 days. More precision and numerical stability can be obtained using high-order integration methods and a smaller step, but this has not been necessary for the moment.

Appendix B presents an estimation of the following input parameters for the model: $\mu_{IH}$ = 3.10 days, $\mu_{AH}$ = 11.36 days, $\mu_{I1}$ = 6.72 days, $\mu_{I2}$ = 13.92 days, $t_D$ = 0.283 and $\gamma$ = 0.165. However, there is no estimation for $P$, $\alpha$, $\omega$ and $\phi_2$. In addition, the date of the first infection is also an unknown (Appendix B gives a plausible range). To solve this problem, an optimization process was launched to minimize the difference between the calculated data and the real one (in a logarithmic scale) for three starting points: beginning with a far date (scenario 1) and ending with a close date (scenario 3). The values of the parameters $P$, $\alpha$, $\omega$ and $\phi_2$ which minimizes the error for each scenario are collected in Table 1. In all scenarios, the restriction of movements imposed by the government has been taken into account by a tenfold reduction of $\omega$, that is, the value of $\omega$ before the day of confinement (March 15th, one day after its official publication) is the one given in Table 1 and it is $\omega/10$ after such day.

|  | **Scenario 1: high $P$** | **Scenario 2:** | **Scenario 3: low $P$** |
| --- | --- | --- | --- |
| **First infection** | **January 25th** | **January 31st** | **February 5th** |
| $P$ (million) | 13.76 | 1.004 | 0.1294 |
| $\alpha$ | 0.501 | 0.502 | 0.538 |
| $\omega$ (pers./pers./day) | 2.70 | 2.76 | 3.32 |
| $\phi_2$ | 0.00342 | 0.0469 | 0.369 |
| Log. Error (total) | 2.07 | 2.07 | 2.06 |
| $R(\infty)$ (million) | 13.64 | 0.9962 | 0.1284 |
| $\phi_D(\infty)P$ | 13203 | 13246 | 13418 |

TABLE 1. Set of parameters found by minimizing the total logarithmic error for three different initial dates. The last two rows give the stationary result for recoveries and deaths.

The errors reached in the three scenarios are almost identical, showing that this error cannot be used to fit the date of the first infection. Indeed, the values of $\phi_2$ and $P$ and this date are strongly correlated, fixing one of them significantly fixes the others. The closer the date of the infection, the larger the value of $\phi_2$. For example, the assumption of March 16th as the initial date leads to $\phi_2 > 1$, which is not possible, and hence this date must be discarded.

In all scenarios the confinement has reduced the impact over the medical services because of the reduction of the susceptible population (note that the reduction of $\omega$ becomes effective after the day the carriers reach its maximum: the value of $\omega$ does not affect much if there are no new susceptible people to be infected). However, scenarios 2 and 3 have a lower number of recoveries than scenario 1 and therefore almost all the Spanish population is susceptible to contagion the day the mobility is present again. If such is the case, an effective practice would be to conduct tests on the maximum number of people in order to reduce its individual mobility as soon as possible.

## V. Discussion

Results show, as expected, that the susceptible population to be exposed is not all the population of Spain. This means that there are regions of Spain (towns, small cities, etc.) which are not accessible to the contagion. Even in cities like Madrid and Barcelona there could be isolated areas with a negligible exposition to the contagion. Obviously the longer the time available for viral spreading, the greater the exposed population. This explains the differences in Table 1. Discovering empirically which scenario is the real one would require conducting tests over a significant part of the population. Without this information we can only make a conjecture about the most plausible scenario. This conjecture comes from comparing the value of $\phi_2$ (note that $\alpha$ and $\omega$ are very similar). In effect, $\phi_2$ changes two orders of magnitude following the change of the susceptible population from 13 to 0.13 million. The risk fraction of 0.00342 is very far away from the values 0.143 and 0.5 estimated as upper bounds in Appendix B whereas the risk factor of 0.369 is in such range. The conclusion is that scenario 3 is more plausible than 1 and that the initial date will probably be between January 31st and February 5th. Assuming, $\phi_2 = 0.167$, the minimization of the error leads to February 3rd, $P = 0.2833$ million, $\alpha = 0.550$, $\omega = 2.95$, $R(\infty) = 0.2811$ million and $\phi_D(\infty)P = 13280$. The solution of the differential equations for this set of input parameters is shown in figure 1, where the matching is significantly good.

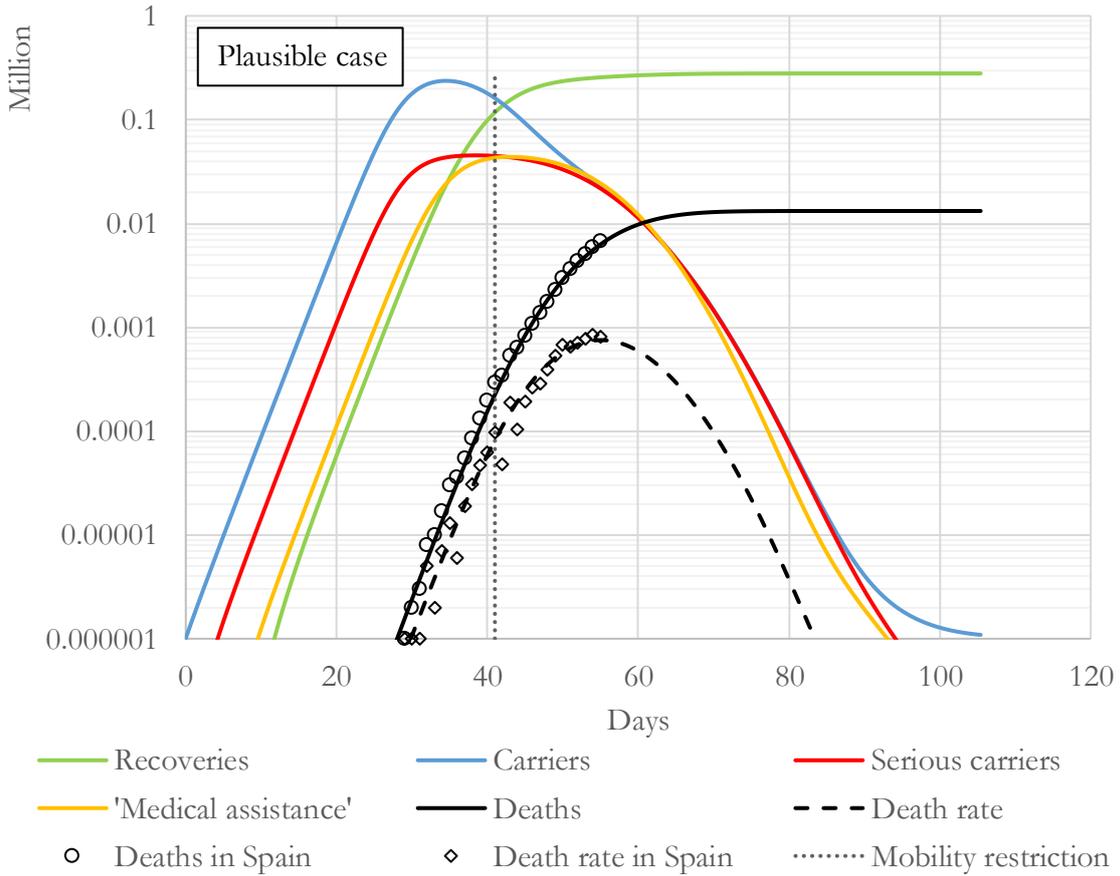

FIGURE 1. Spreading of CoVid-19 in Spain assuming February 3rd as the initial day. Solid lines are calculated using the CIR model with the input parameters: February 3rd, $P = 0.2833$ million, $\alpha = 0.550$ and $\omega = 2.95$. Real data comes from reference [1].

Note that the real active cases for mild and serious cases has not been used to fit any parameter since their definitions are not clear (are the infected who stay at home counted as mild cases?) and probably does not coincides with the ones given in this article. At the moment of writing this article, the number of active cases in Spain was 80110 [1] whereas the final number calculated for Figure 1 is 281100 but can be much higher if the initial date is moved far away. For this reason, only deaths have been used for fitting purposes: the uncertainty on reported deaths, although not zero, is lower. To illustrate this fact, Figure 2 shows the calculated curves for the case less probable (i.e., scenario 1 in Table 1). As can be seen, the matching is as good as the one obtained for the case of figure 1. However, the difference in the number of final recoveries is huge: 13 million (near 27% of the Spanish population) in the less probable case and 0.2811 million (near 0.6%) in the other case.

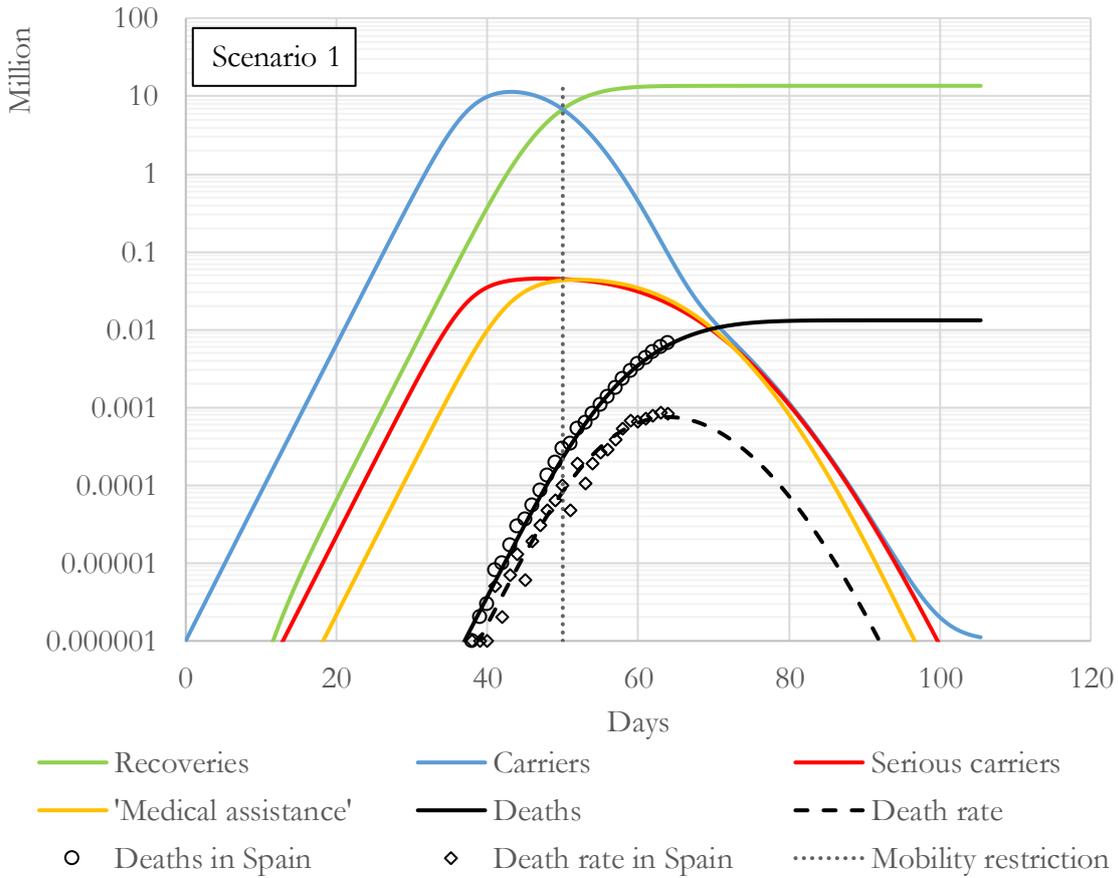

FIGURE 2. Spreading of CoVid-19 assuming the less probable case. Solid lines are calculated using the CIR model with the input parameters shown in Table 1 (scenario 1). Real data comes from reference [1].

## VI. Conclusion

A new predictive model based on differential equations and convolutions has been described and used to estimate the death rate by CoVid-19 in Spain. It has shown that 1) only one set of parameters is required to obtain a prediction over a full curve; 2) there is a strong dependence between the date of the first infection and the susceptible population and risk fraction, fact which allows the estimation of such day; 3) it can be used to estimate the number of susceptible people in near-future massive infections; and 4) it can be used to estimate the demand for hospital services and the effect of different governmental actions.

## Appendix A. Stochastic model

Let $h(\Delta t, \mu)dt$ be the probability that an infected individual has of suffering a given event (for example, developing symptoms, leaving the UCI, leaving the hospital, recuperating or passing away, etc.) just in time $t$ after having being infected during a time $\Delta t$. It is plausible to assume that it responds to a general distribution of the form (note that it could be substituted by any other distribution without changing the model):

$$h(\Delta t, \mu) = \frac{m/\mu}{\Gamma\left(\frac{n+1}{m}\right)} \left(\frac{\Delta t}{\mu}\right)^n e^{-\left(\frac{\Delta t}{\mu}\right)^m}$$

with

$$\int_0^\infty x^n e^{-x^m} dx = \frac{\Gamma\left(\frac{n+1}{m}\right)}{m}$$

It is convenient to use a new dimensionless function $s$ and a new dimensionless independent variable defined as follows

$$h(\Delta t, \mu)\mu = h(x\mu, \mu)\mu = \frac{m}{\Gamma\left(\frac{n+1}{m}\right)} x^n e^{-x^m} = s(x; n, m)$$

Function $s(x)$ has a maximum at

$$\frac{ds}{dx} = 0 \Rightarrow x_M = \left(\frac{n}{m}\right)^{1/m}$$

Function $h(\Delta t, \mu) = \frac{s(x;n,m)}{\mu} = \frac{x\, s(x;n,m)}{\Delta t}$ has a maximum with $\mu$ at

$$\frac{d(xs)}{dx} = 0 \Rightarrow x_m = \left(\frac{n+1}{m}\right)^{1/m}$$

Note that this value is very interesting because is the Maximum Likelihood Estimator for $\mu$ knowing that only one event has happened at $\Delta t$.

The mean $\langle \Delta t \rangle$ and the standard deviation $\sigma$ are respectively

$$\frac{\langle \Delta t \rangle}{\mu} = \int_0^\infty x\, s(x)\, dx = \frac{\Gamma\left(\frac{n+2}{m}\right)}{\Gamma\left(\frac{n+1}{m}\right)}$$

$$\frac{\sigma^2}{\mu^2} = \frac{\langle \Delta t - \langle \Delta t \rangle \rangle^2}{\mu^2} = \int_0^\infty (x-1)^2 s(x)dx = \frac{\Gamma\left(\frac{n+3}{m}\right) - 2\Gamma\left(\frac{n+2}{m}\right) + \Gamma\left(\frac{n+1}{m}\right)}{\Gamma\left(\frac{n+1}{m}\right)}$$

For $m = 2$, $\frac{\sigma^2}{\mu^2}$ has a minimum at $n = 1.333839051$, what leads to $\frac{\sigma^2}{\mu^2} = 0.2205062985$.

### Particular case

A Gaussian-like distribution has $m = 2$. A stochastic distribution with enough uncertainty has $\sigma^2 = \mu^2$, so that, $n = 6.484478437$, $\langle \Delta t \rangle = 1.871119609\mu$, $x_M = 1.800621898$ and $x_m = 1.934486810$.

The first quartile is at $\frac{m}{\Gamma\left(\frac{n+1}{m}\right)} \int_0^{x_1} x^n e^{-x^m} dx = 0.25 \Rightarrow x_1 = 1.524527018$.

The median is at $\frac{m}{\Gamma\left(\frac{n+1}{m}\right)} \int_0^{x_2} x^n e^{-x^m} dx = 0.5 \Rightarrow x_2 = 1.847892050$

The third quartile is at $\frac{m}{\Gamma\left(\frac{n+1}{m}\right)} \int_0^{x_3} x^n e^{-x^m} dx = 0.75 \Rightarrow x_3 = 2.192178048$

The dimensionless IRQ is $\varepsilon = \frac{x_3 - x_1}{x_2} = 0.3613041303$.

The probability of 97.5% is at $\frac{m}{\Gamma\left(\frac{n+1}{m}\right)} \int_0^{x_c} x^n e^{-x^m} dx = 0.975 \Rightarrow x_c = 2.894362301$.

**Model of recovery**

$$h(\Delta t, \mu_R) = \frac{0.4563477340}{\mu_R} \left(\frac{\Delta t}{\mu_R}\right)^{6.484478437} e^{-\left(\frac{\Delta t}{\mu_R}\right)^2}$$

**Model of input**

$$h(\Delta t, \mu_{IH}) = \frac{0.4563477340}{\mu_{IH}} \left(\frac{\Delta t}{\mu_{IH}}\right)^{6.484478437} e^{-\left(\frac{\Delta t}{\mu_{IH}}\right)^2}$$

**Model of output**

$$h(\Delta t, \mu_{AH}) = \frac{0.4563477340}{\mu_{AH}} \left(\frac{\Delta t}{\mu_{AH}}\right)^{6.484478437} e^{-\left(\frac{\Delta t}{\mu_{AH}}\right)^2}$$

## Appendix B.

Reference [2] reports statistics over 191 patients, of whom 137 were discharged and 54 died in hospital. This means that 28.3% of patients died (26% required ICU), which leads to $t_D = 0.283$. We use this data to obtain the following parameters for the serious infection, that is, for the type 2.

|  | $x_1 \mu$ | $x_2 \mu$ | $x_3 \mu$ | Estimation | $\frac{x_3 - x_1}{x_2}$ | Deviation from $\varepsilon$ |
|---|---|---|---|---|---|---|
| Time from illness onset to ICU admission (days) | 8 | 12 | 15 | $\mu_{IICU} - \mu_{IH} = \frac{12}{x_2} = 6.49$ | 0.58 | 61% |
| Time from illness onset to death or discharge (days) | 17 | 21 | 25 | $\mu_{OH} = \frac{21}{x_2} = 11.36$ | 0.38 | 5.6% |
| Duration of viral shedding after CoVid-19 onset (days) | 16 | 20 | 23 | $\mu_{I2} - \mu_{IH} = \frac{20}{x_2} = 10.82$ | 0.35 | 2.8% |

Reference [1] reports 81400 cases in China of whom 3304 are deaths, thus, using the value $t_D = 0.283$, the risk fraction is roughly estimated as $\phi_r \approx 3304/0.283/81400 = 0.143$. However, this value should be treat as a upper limit since we do not know if there were cases not diagnosed.

Reference [3] reports that a Spanish woman who was infected on February 29[th] began symptoms on March 5th, that is, 5 or 6 days after the infection. She did recover completely on March 12[th], that is, 12 or 13 days after the infection. The CoVid-19 test was positive on March 10[th]. She did not transmit the virus to any of her relatives nor to any of the 11 people who she met before knowing she was infected. We can use this case to estimate the parameters for mild infections, that is, for type 1: $\mu_{I1} = \frac{13}{x_m} = 6.72$ days. We can also use this case to estimate the parameter for developing symptoms and entering the medical service: $\mu_{IH} = \frac{6}{x_m} = 3.10$ days. This leads to a mean value of 5.8 days, which is very similar to those (3.0 days, 5.2 days or 6.4 days) reported by [1] depending on the group of people studied.

Reference [4] reports that a party held at the end of February in Spain brought together 80 (apparently healthy) people. The result were 14 infections, of whom 7 were hospitalized and 1 was admitted in the ICU. He was admitted in the ICU on March 10th, nearly 12 days after the party, this means that for type 2, $\mu_{IICU} = \frac{12}{x_m} = 6.20$ days, which is a result very similar to that obtained previously from reference [3]. It is relevant to remark that this person was healthy, athletic, non-smoking and in his 50s, thus we conclude that all the population is susceptible to be in the type 2; however, only 7 out of 14 were seriously affected and 1 out of 14 was very seriously affected. This sets a rough estimation for the fraction of type 2 infections, $\phi_2 = \phi_r \approx 0.5$, but note that this number should be taken as a superior limit because there could be non-detected carriers. In addition, it seems plausible that, at that party, there was only one initial carrier of type 1 and hence, the success in the contagion can be estimated as $\gamma = 13/79 = 16.5\%$.

Reference [5] reports that the first case in Spain (in Gomera island) was confirmed on January 31st, 2020. The infected person had mild symptoms and was discharged on February 14th. The contagion came from a German person who had been diagnosed. This case and many others like the one reported in reference [3] make very plausible the hypothesis of having two types of infection: mild and serious. The first case on the Iberian Peninsula (in Catalonia) was a 36-year-old woman who was confirmed on February 26th. Supposedly, she was exposed to the virus in the north of Italy (Milan and Bergamo) between the days 12th and 22nd of February. Taking into account an incubation period of 6 days, we could have two rough estimations for the initial date: January 25th and February 23rd. Almost for sure, there were other infections which were not detected because they laid in the mild-condition group (note that this group includes even those infected who had absence of symptoms) and hence the initial infection could have happened, as a first guess, near February 5th. This data has great uncertainty and must be the object of further discussion.

## Appendix C

Assuming that the fraction of carriers is small and that there is only one type of infections, equations (11) and (12) leads to

$$\frac{d\phi_I(t)}{dt} \cong (1-\alpha)\phi_C(t)\omega\gamma$$

$$\frac{d\phi_C(t)}{dt} = \frac{d\phi_I(t)}{dt} - \int_0^t \frac{d\phi_I(t-\Delta t)}{dt} h(\Delta t, \mu_R) d\Delta t$$

For long times $t \gg \Delta t$, it happens that $\phi_C(t - \Delta t) \xrightarrow{t \gg \Delta t} \phi_C(t) - \frac{d\phi_C(t)}{dt}\Delta t$. When $\Delta t \gg \mu_R$, the distribution tends to zero, $h(\Delta t, \mu_R) \xrightarrow{\Delta t \gg \mu_R} 0$, and hence $\int_0^t \phi_C(t - \Delta t) h(\Delta t, \mu_I) d\Delta t \xrightarrow{t \gg \mu_I} \phi_C(t) \int_0^\infty h(\Delta t, \mu_R) d\Delta t - \frac{d\phi_C(t)}{dt} \int_0^\infty \Delta t\, h(\Delta t, \mu_R) d\Delta t = \phi_C(t) - \frac{d\phi_C(t)}{dt}\langle\Delta t\rangle$. Thus, the differential equation becomes $(1 - (1-\alpha)\langle\Delta t\rangle\omega\gamma)\frac{d\phi_C(t)}{dt} \cong 0$. As long as $\omega\gamma$ and $\langle\Delta t\rangle$ are independent parameters, $(1-\alpha)\langle\Delta t\rangle\omega\gamma \neq 1$ holds in general, and hence $\frac{d\phi_C(t)}{dt} \cong 0$. That is, for long times, $\phi_C$ tends to be constant.

# References

[1] www.worldometers.info/coronavirus/countries/spain/, visited on March 29th, 2020.

[2] Fei Zhou*, Ting Yu*, Ronghui Du*, Guohui Fan*, Ying Liu*, Zhibo Liu*, Jie Xiang*, Yeming Wang, Bin Song, Xiaoying Gu, Lulu Guan, Yuan Wei, Hui Li, Xudong Wu, Jiuyang Xu, Shengjin Tu, Yi Zhang, Hua Chen, Bin Cao, "Clinical course and risk factors for mortality of adult inpatients with COVID-19 in Wuhan, China: a retrospective cohort study", *www.thelancet.com,* Published online March 9, 2020 https://doi.org/10.1016/S0140-6736(20)30566-3.

[3] María Traspaderne (EFE), "Alejandra, 20 años, curada del Covid-19: No me sentía mal y seguía quedando", *Heraldo* (published online), March 3rd, 2020.

[4] Pedro Ybarra Bores, "Ingresados cuatro miembros de una misma familia sevillana por coronavirus, uno de ellos en UCI", *ABCdesevilla* (published online), March 17th, 2020.

[5] "¿Cuál fue el primer caso de coronavirus en España y en la península?", *20minutos* (published online), March 13th, 2020.